\documentclass[]{aa}

\usepackage{graphicx}
\usepackage{subfigure}

\begin{document}


\title{Infrared emission towards \object{SN~1987A} 11 years after outburst:
Measurements with ISOCAM
\thanks{Based on observations with {\em ISO}, an {\em ESA} project with instruments funded by {\em ESA} 
member states (especially the P/I countries France, Germany, the Netherlands and the United Kingdom)
with participation of {\em ISAS} and {\em NASA}.}
}
\author{J\"org Fischera \and Richard J. Tuffs \and Heinrich J. V\"olk}

\offprints{J\"org Fischera, \email{fischera@mickey.mpi-hd.mpg.de}}

\institute{Max-Planck-Institut f\"ur Kernphysik, Saupfercheckweg 1, 69115 Heidelberg}

\date{Received  ---/ Accepted ---}

\authorrunning{Fischera et al.}
\titlerunning{Infrared emission towards \object{SN~1987A} 11 years after outburst}

\abstract{
We present measurements of the mid-infrared (MIR) emission from SN~1987A, 
made using the Infrared Space Observatory (ISO) 11 years 
after outburst. They are the only late epoch detections 
of this source in the thermal IR regime.
The position of the source, determined from an offset to an IR-emitting star, 
suggests that the emission is associated with SN~1987A or its 
extended supernova remnant (SNR).
A predominantly circumstellar origin is however suggested by
the size and orientation of the IR-emitting region, which is 
comparable with the extension of the inner ring seen with the 
Hubble Space Telescope (HST). The emission is most probably from 
collisionally-heated circumstellar grains embedded in
shocked gas downstream of the blast wave. The MIR extent is consistent with
the hypothesis that the blast wave was propagating into material of
moderate density interior to the thick inner ring at the epoch of the ISOCAM observations.
}

\maketitle


\section{Introduction}

Supernovae are thought to play an important role in 
the life cycle of interstellar dust grains. 
On the one hand, 
shocks driven by supernovae into the interstellar medium (ISM)
are considered to be responsible
for the destruction of interstellar grains 
(see e.g. Shull, \cite{Shull77}, Draine \& Salpeter, \cite{Draine79b}, 
Tielens et al., \cite{Tielens94}, Jones et al., \cite{Jones94}). 
On the other hand, supernovae have also been proposed as
sources of interstellar dust (e.g. Hoyle \& Wickramasinghe, 
\cite{Hoyle70}, Dwek \& Scalo, \cite{Dwek80}). 


SN~1987A in the Large Magellanic Cloud (LMC) was the first supernova in which 
the condensation of dust in the central metal-rich 
part of the expanding ejecta has been observed. 
The appearance of a continuum emission in the 
IR in the spectrum of SN~1987A 
614 days after outburst is attributed to grains formed 
after at least 500 days (e.g. Wooden, \cite{Wooden97}). 
The infrared emission from the condensates was
(maybe except at the onset of dust formation; 
Roche et al., \cite{Roche93}) optically thick at the early epochs  when the
IR-emission was still bright enough to be observable from ground.
Therefore, it was not possible 
to get clear information about their properties and in particular their mass.

Observations with the HST have revealed a bipolar ring structure 
consisting of a thick inner ring centred at the position
of the supernova, and two thin outer rings of 
larger radius at each side (Burrows, \cite{Burrows95}). 
Radio and X-ray observations suggest that, after an initial period
of free expansion, the blast wave started to propagate through 
a region of moderate density with 
$n_{\rm H}\approx 100~{\rm cm^{-3}}$ 
interior to the thick inner ring 
(Chevalier \& Dwarkadas, \cite{Chevalier95}). 
Chevalier \& Dwarkadas suggested this region to be an HII region, illuminated by
the blue supergiant supernova progenitor and composed of
wind material ejected during the prior red supergiant phase.

If the HII region contains dust,
extended thermal emission 
from collisionally heated grains embedded in the shocked gas 
downstream of the blast wave is to be expected.
The presence of circumstellar grains was already invoked at early epochs
to account for observations of a prompt echo effect in the mid-infrared 
(Rank et al., \cite{Rank88}, Chabalev et al., \cite{Chabalev89}). 
It has also been suggested that circumstellar grains can account for observations after 580~days
showing an extension of $1\farcs 5$ at $9~\mu{\rm m}$ 
(Roche et al., \cite{Roche93}). Light echos in the optical have 
further been used to reconstruct a three dimensional model of the larger
scale structures around SN~1987A (Crotts et al., \cite{Crotts95}).


The ISO mission (Kessler et al., \cite{Kessler96}) gave a unique opportunity to probe 
the early evolution of SN~1987A in the IR prior to the 
envelopment of the thick inner ring by the blast wave.
In this paper we present sensitive
measurements made with the ISOCAM instrument (Cesarsky et al., \cite{Cesarsky96}) on board ISO
11 years after outburst. Thus far, these measurements constitute
the only late epoch 
detection of the source in the IR regime. 
Preliminary results
were presented by Tuffs et al. (\cite{Tuffs98}).
Here we demonstrate that the detected IR emission is  
most probably of circumstellar origin. 

The paper is structured as follows:
In Sect. 2 we present the ISOCAM measurements and  
describe the reduction
and calibration of the data.
In Sect. 3 we derive the flux density, position, 
orientation and size of the mid-infrared (MIR) source, showing that the
IR emission from the remnant is resolved, with an extent 
comparable to that of the thick inner ring.
These results are discussed in Sect. 4.
A summary of the paper is given in Sect. 5.

Throughout the paper we adopt a distance of 51 kpc to the supernova 
as was used to analyse the X-ray observations
by Hasinger et al. (\cite{Hasinger96}). This is close to the distance of $51.2\pm3.1$ kpc
derived by Panagia et al. (\cite{Panagia91}) but slightly larger than the
values given in later publications
(Gould, \cite{Gould95}, Sonneborn et al., \cite{Sonneborn97}, Gould, \cite{Gould98}).

\section{Observations}

The observations were made with the ISOCAM instrument 
in the $LW2$, $LW10$ and $LW3$ broad-band filters, whose pass bands are
centred on 6.75, 12 and $14.3~\mu{\rm m}$ respectively. 
An observation log summarizing
the principal observational parameters is given in table \ref{observ_log}.

The measurements in $LW10$ and $LW3$ were optimized for obtaining
sensitive photometry and structural information of a faint compact target.
A pixel field of view (PFOV) 
of $3\arcsec $ was chosen as for the anticipated backgrounds
it gave the best trade off between angular resolution and 
detector illumination. We used a $3\times 3$ raster with raster interval 
$10\arcsec\times 10\arcsec$ to sample the sky at intervals of $1\arcsec$.
This comfortably oversampled the point spread function ($psf$), which has a 
$FWHM$ of $\sim 4\arcsec$ at $10~\mu{\rm m}$ for the $3\arcsec$ PFOV.

The raster interval of $10\arcsec$ was the minimum
interval that provides a sky sampling for 
which each view of the target in a given pixel 
is immediately preceded or followed by a 
view of the underlying background. 
This optimized the knowledge of the relative
response of the detector between source and background,
which in turn could be used to minimize the effect of residual 
uncertainties in the flat--field response of the $32 \times 32$ pixel detector 
on the determination of source structure.
The observations were separated in epoch over the period September 1997 to
February 1998. The measurement in the $12~\mu{\rm m}$ band 
was a follow up observation to the detection in the $14.3~\mu{\rm m}$,
while the measurement in the $6.75~\mu{\rm m}$ band was part of an independent
program of large-scale mapping of the LMC led by L. Vigroux.


\begin{table}[htbp]
 \begin{minipage}[t]{\hsize}
  \caption{Observation log}
  \label{observ_log}
  \begin{tabular}{l||c|c|c}
  Filter & $LW2$ & $LW10$ & $LW3$ \\
  \hline
  TDT & 667017 & 811022 & 750019 \\
  observer & L. Vigroux & R.J. Tuffs & R.J. Tuffs \\
  $\lambda_{\rm ref} ~[\mu{\rm m}]$ & 6.75 & 12. & 14.3 \\
  filter range $[\mu{\rm m}]$ & 5.00-8.50 & 8.00-15.00 & 12.0-18.\\
  observation date & 13.9.97 & 5.12.97 & 3.2.98  \\
  day after outburst\footnote{The time of the outburst is taken to be February 23.316, 1987 (Arnett et al., \cite{Arnett89}).} & 3855.2 & 3998.7 & 3937.7 \\
  raster & $2\times 2$ &  $3\times 3$ & $3\times 3$ \\
  pixel size & $6\arcsec$  & $3\arcsec$ & $3\arcsec$ \\
  step size & $24\arcsec$ & $10\arcsec$ & $10\arcsec$ \\
  read out interval $t_{\rm int}$ & 2.10 s & 5.04 s & 5.04 s \\
  reads per pointing $n_{\rm i}$ & 31 & 19 & 19 \\
  \end{tabular}
 \end{minipage}
\end{table}

\subsection{Calibration and Data reduction}

The data were first corrected for dark current 
using the model of the \emph{Cam Interactive Analysis} 
(CIA) package (Ott et al., \cite{Ott98}; version April 2000).
This was followed by glitch removal using the 
\emph{Multi resolution Median Transform}, also provided by CIA.
The data were then inspected and residual events and
strong longer-lived glitch-induced distortions were removed manually.
To allow an accurate determination of
the illumination seen by each single pixel during the raster, the  
data were then corrected for the transient response behavior
of the detector. To this end we developed and implemented a routine
based on the drift model of the ISOPHOT-S 
detector on board ISO (Schubert, \cite{Schubert95}). The routine
derived the time constants and jump factors
determining the drift behavior of the individual detector pixels 
from the measured data.

\begin{figure*}[htbp]
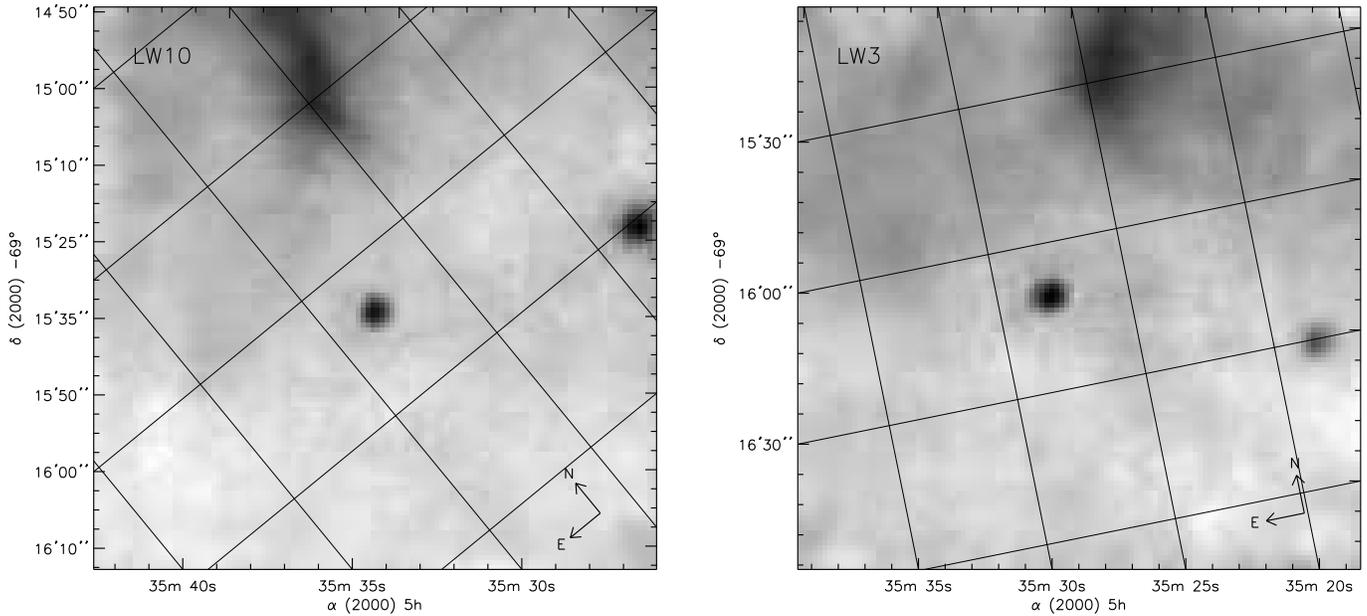

  \resizebox{0.48\hsize}{!}{\includegraphics{ms1827f1.epsi}}      \hfill
  \resizebox{0.48\hsize}{!}{\includegraphics{ms1827f2.epsi}}  
  \caption{
    \label{ISOCAM-fig}
    Region around the SN~1987A in the MIR, as seen at $12~\mu{\rm m}$ ($LW10$) and 
    $14.3~\mu{\rm m}$ ($LW3$).
    The brightness increases with darkness and ranges from 0.215 to $0.306~{\rm mJy/[\arcsec]}^2$ ($LW10$)
    or from 0.356 to $0.426~{\rm mJy/[\arcsec]}^2$ ($LW3$) respectively.
    The source close to the middle is emission from the remnant of SN~1987A.
    The IR-emission at the border of the image is the reference star, that we used to
    derive the absolute coordinates of the IR-emission from the remnant of SN~1987A.
    The bright emission in the north is an arm of 30 Doradus. The two individual sources
    show different colours, suggesting a lower temperature for the IR-emission
    from SN~1987A than for the star.}
\end{figure*}

To create a map showing the background structure underlying the sources
we applied a flat--field correction, which was optimized for the smooth
variation of the background. The flat--field was derived from a map, which had already been corrected
using a library flat provided by CIA and which had been spatial filtered to reduce the noise.
The flat--field--corrected data were then binned (without interpolation)
on a grid with a resolution of $1\arcsec$.
For visualization the data were further filtered by setting pixel values differing more than one sigma 
from their surroundings to the average value of the surrounding pixels (the filter was not
applied for obvious discrete sources).
Then map pixels without any data were set to the average values of the neighbouring pixels.
Finally, we convolved the image with a Gauss function with a variance of $0\farcs6$.
The resulting maps are shown in \mbox{Fig. \ref{ISOCAM-fig}}.

The quantitative analysis of
the individual compact sources seen in the ISOCAM maps 
was done directly from the observed brightnesses
as a function of sky position, without use of maps, as 
described in Sect. \ref{sectionparameter}. This required a
different calibration procedure compared to that for the maps.
For the observations at 12 and $14.3~\mu{\rm m}$ we made use of the 
almost flat emission around
the two sources and normalized all pixel values at individual raster 
positions
with the corresponding values of the 
background. This was done by making a linear fit to the variation 
of (responsivity drift-corrected) illumination in each given detector
pixel with spacecraft raster position, leaving out raster positions
where the pixel was viewing the source.
In the case of the measurement at $6.75~\mu{\rm m}$,
we used the library flat
provided in CIA, as the raster pattern was not suitable for this method.
For the measurement at 12 and at $14.3~\mu{\rm m}$
the sky coordinates of each detector pixel at each raster pointing were
evaluated from the actual (post posteri) pointing position of the satellite 
using the median of all the data used. 
To derive the pixel coordinates of the observation at $6.75~\mu{\rm m}$ we
made use of the astronomical information provided by the CIA-program.
Corrections were applied
for the astrometrical distortions in the ISOCAM field 
(Aussel, \cite{Aussel98}) induced by the field lens using the polynomial
correction coefficients for the different filters as given in CIA.

\subsection{The ISOCAM maps}

\label{isocammaps}
The final images show two clear individual sources. The source close 
to the middle corresponds to
emission from the remnant of SN~1987A, whereas the source near the edge 
is an IR-emitting star. 
The two sources show different 
colours, indicating a higher temperature for the star.
The serendipitous 
detection of the star provided a unique possibility to
determine an accurate position for the central
MIR source near SN~1987A (see Sect. \ref{coordinatesection}).
The bright ridge of emission
to the NNW of SN~1987A is an arm of the 30 Doradus nebula. 
Although backgrounds are dominated by smooth emission from zodiacal 
light, there is some evidence for faint cirrus structure at the 1\% level
in the southern part of the field.
At both wavelengths the IR emission near the map centre is offset several
arcsec from the nominal position of SN~1987A.
These offsets are partly due to the inaccuracy of the position of the
satellite but are mainly attributable to a displacement of the field lens of the camera 
(Blommaert \& Cesarsky, \cite{Blommaert98b}).

\section{Parameters of the discrete sources}

\label{sectionparameter}
\begin{figure*}
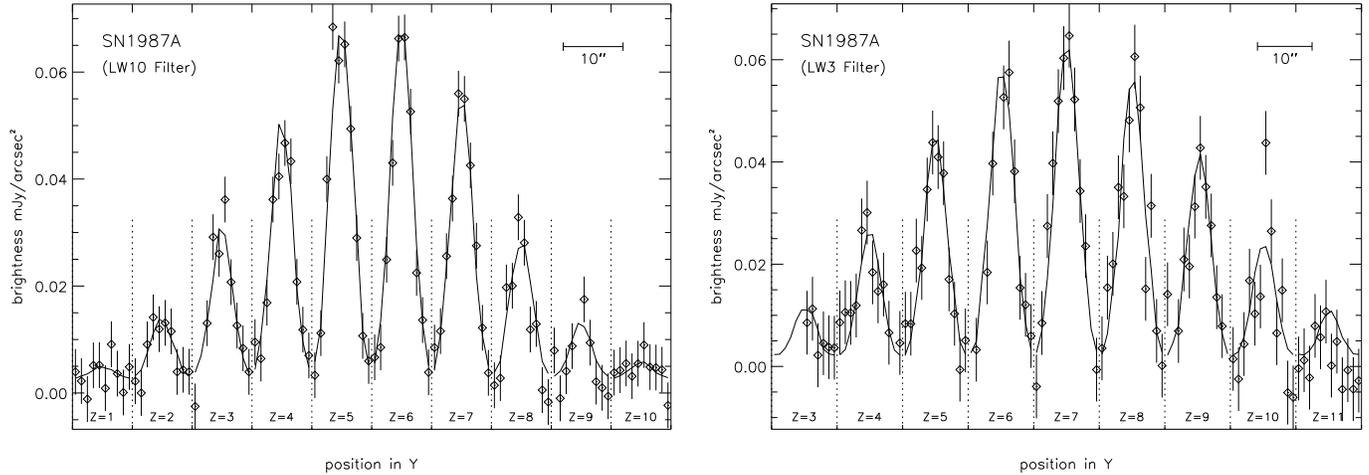

  \resizebox{0.49\hsize}{!}{\includegraphics{ms1827f3.epsi}}
  \hfill
  \resizebox{0.484\hsize}{!}{\includegraphics{ms1827f4.epsi}}
  \caption{
    \label{fitresult}
    Best fits of the theoretical image (solid line) to the measured brightnesses (diamonds) 
    of the remnant of SN~1987A
    at $12~\mu{\rm m}$ ($LW10$) and $14.3~\mu{\rm m}$ ($LW3$). 
    They are shown as scans
    along lines (at fixed $Z$ in the coordinate system of the satellite)
    through a subimage centred on SN~1987A. The dotted vertical lines mark the
    beginning and the end of the scans.
    To calculate the $psf$ for each filter the spectral shape was assumed to be
    $F_\lambda\propto\lambda^{-k}$ with a spectral index $k=1.08$. The $FWHM$ of the 
    circular Gaussian source is $1\farcs 50$ ($LW10$) and $1\farcs 13$ ($LW3$).
    }
\end{figure*}

Various parameters of the discrete sources, expressed as the vector $\vec a$,  were derived
using a non-linear $\chi^2$-fit, where we compared the measured data $f(\vec x_{i})$
and uncertainties $\sigma_{i}$ 
at the sky positions $\vec x_{i}$ around the individual sources
with a model image $B(\vec x_{i},\vec a)$:
\begin{equation}
  \label{chisqr_equation}
  \chi^2(\vec a)=\sum_{i}\frac{
    \left(B(\vec x_{i},\vec a)-f(\vec x_{i})\right)^2}{\sigma_{i}^2}.
\end{equation}
The model image $B(\vec x,\vec a)$ is in general the convolution of the theoretical $psf$
corresponding to the telescope optics and the spectral response of the used 
filter band pass  
with the source function (taken to be a Gaussian), 
on a certain background. For a symmetrical Gaussian fit $\vec a$ is 
$[F_{\rm Q}, \vec x_0, \sigma_{\rm Q}, I_U]$ and the model image given by:
\begin{eqnarray}
  \label{imagefkt}
  B(\vec x,\vec a) &=&
  I_{\rm U}+\int J_{\rm F}(\vec x-\vec x')\nonumber\\
  &&\times \frac{F_{\rm Q}}{2 \pi \sigma_{\rm Q}^2} \exp\left({{-\frac{\left(\vec x'-\vec x_0\right)^2}{2\,\sigma_{\rm Q}^2}}}\right)\,{\rm d}\vec x',
\end{eqnarray}
where $J_{\rm F}(\vec x)$ is the theoretical broad-band $psf$ (see appendix \ref{psfsection}), 
$F_{\rm Q}$ the flux density, $\vec x_0$ the coordinates, $\sigma_{\rm Q}$ the variance of the
Gaussian source and $I_U$ the background. 
The expression \ref{chisqr_equation} was minimized to derive $F_{\rm Q}$,
$\vec x_0$, $I_U$ (at $6.75~\mu{\rm m}$ only) and $\sigma_{\rm Q}$ (only at $12$ and $14.3~\mu{\rm m}$). 

To achieve a higher accuracy, in 
particular for the determination of the size of the emitting region, 
we took into account the dependence of beam profile 
on source colour. This was done by 
averaging simulated beam shapes obtained at a series of 
wavelengths across the broad filter pass bands,
weighted according to a power law 
source spectrum $F_{\lambda}\lambda^k=const.$
with a spectral index $k$ (see appendix \ref{psfsection}).
For the star the spectral index was always taken 
to be $k=4$ appropriate to a Rayleigh-Jeans spectrum.
For the source associated with the supernova $k$ was estimated from the finally determined
colour corrected (see e.g. Blommaert, \cite{Blommaert98a}) flux densities to be $1.08\pm 0.32$.
The model image was calculated on a grid 
with a resolution of $1/3\arcsec$. In the fitting procedure
the predicted fluxes at the observed positions 
were found using linear interpolation.

The true uncertainties $\sigma_{i}$
in each individual measurement for a detector pixel at a given
spacecraft raster pointing direction are 
difficult to estimate and may be partly
affected by systematic uncertainties.
This particularly applies to measurements viewing 
the discrete sources, where 
the illumination and detector response function may vary with position 
over the field of view of the individual detector pixels.
We therefore assumed in nearly all model calculations  
that the uncertainties of all measurements are equal. 
This is justified by the fact that the distribution of the residuals was 
found to be normal.
The only exception to this was made for the 
measurement of the star at $6.75~\mu{\rm m}$, where the uncertainties of
the high source fluxes were mainly attributed to a
particularly strong transient behavior.

The absolute uncertainties $\sigma_i$ of the measured data 
were chosen such that the value of the minimum $\chi^2$ (Eq.~\ref{chisqr_equation})
corresponds to a confidence level of 90\%.
In the case of the 
measurements at 12 and at $14.3~\mu{\rm m}$ 
we made a special fit letting $F_{\rm Q}$, $\vec x_0$, $\sigma_Q$ and $k$ as free variables.
The joint uncertainties of $\nu$ parameters were estimated 
by finding the range within which the function $\Delta\chi^2_{\nu}=\chi^2_{\nu}-\chi^2_{\rm min}$
did not exceed the value corresponding to a confidence level of one sigma,
letting all other parameters as free variables.
Thus, apart from the jointly determined uncertainties in the coordinates, which were estimated from $\Delta\chi^2_2=2.30$,
the uncertainties in individual parameters correspond to $\Delta\chi^2_{1}=1$ (see e.g. Press et al., \cite{Press92}). 

\subsection{Fits with a symmetrical Gaussian as IR-source}

\label{basicfitresults}
In the case of the measurement at $6.75~\mu{\rm m}$ the parameters varied were
$F_{\rm Q}$,
the coordinates $\vec x_0$ and the background brightness $I_{\rm U}$. 
We used a model image corresponding
to an ideal point source ($\sigma_{\rm Q}=0$) 
assuming a spectrum with $k=1$ for 
the IR emission from the supernova remnant.

At 12 and at $14.3~\mu{\rm m}$ the flux density and the coordinates were optimized together
with the extension of the source. The variables varied were therefore
$F_{\rm Q}$, $\vec x_0$ and $\sigma_{\rm Q}$. 
Because of the normalization of the pixel data to the background we held $I_{\rm U}$ fixed.
The uncertainties in $F_{\rm Q}$ and $\vec x_0$ were estimated for a fixed (most probable) source size.
The best fits to the measured data with $LW10$ and $LW3$ are shown in Fig.~\ref{fitresult}.

\begin{table*}[htbp]
 \begin{minipage}[t]{\hsize}
  \setlength{\tabcolsep}{1.2mm}
  \caption{
    \label{fitparam}
    Spherical Gauss fits to the IR emission associated with SN~1987A and the reference star}
  \begin{tabular}{l||l|l|l||l|l|l}
   \hline
   \hline
    & \multicolumn{3}{c||}{SN~1987A}    & \multicolumn{3}{c}{reference star}\\
    \hline
    Filter & $LW2$ & $LW10$ & $LW3$ & $LW2$ & $LW10$ & $LW3$\\
    $\lambda_{\rm ref}~[\mu{\rm m}]$ & 6.75 & 12. & 14.3 & 6.75 & 12. & 14.3 \\
    ${\rm ADU\footnote{Analog to Digital Units} /(mJy/pixel)}$\footnote{Siebenmorgen et al., \cite{Siebenmorgen99}.}
 & $2.32\pm3.3\%$ & $4.23\pm3.9\%$ & $1.96\pm4.8\%$ & $2.32\pm3.3\%$ & $4.23\pm3.9\%$ & $1.96\pm4.8\%$ \\
    beam size ($FWHM$ $[\arcsec ]$)\footnote{Average $FWHM$ of cuts in Z and Y through the maximum of the theoretical $psf$. \\
      Uncertainties correspond to the difference of these two values.} & $6.887\pm 0.002$ & $4.195\pm 0.025$ & $5.109\pm0.009$ & $6.222\pm 0.001$& $3.945\pm 0.026$ & $4.955\pm 0.008$  \\
    \hline    
    spectral index $k$\footnote{Spectral index assumed in calculating the theoretical $psf$ (see text).} 
       & 1. & 1.08 & 1.08 & 4 & 4 &4 \\
       derived source size ($FWHM$ $[\arcsec ]$)\footnote{$FWHM$ of the symmetrical Gaussian model assumed for the source. Uncertainties
         are given as 1 $\sigma$.} 
       & -- & $1.50_{-0.24}^{+0.21}$ & $<2.3\,(3\,\sigma)$ & -- & $2.26_{-0.30}^{+0.27}$ & $<2.4\,(3\,\sigma)$ \\
    $\chi^2_{\nu}~(\nu)$ & 0.76 (51) & 0.82 (94) & 0.83 (127) & 0.77 (60) & 0.69 (41) & 0.78 (75) \\
    Uncertainty of pixel values $\sigma/I_{U}$ & 3.6\% & 1.9\% & 1.7\% & 12.\%\footnote{Median of the uncertainties.} 
            & 2.5\% & 1.5\% \\
    $F_{\mathrm{Q}}/I_{\mathrm{U}}~\mathrm{[[\arcsec ]^2]}$ & $6.54\pm 25\%$ & 
        $10.62\pm 1.8\%$ & $7.608\pm 2.6\%$ & $116.2\pm 5.5\%$ & 
        $16.11\pm 3.1\%$ & $4.879\pm 5.0\%$ \\
    background $I_{\mathrm{U}}~\mathrm{[mJy/[\arcsec]^2]}$ & $0.0759\pm 3.3\%$ & 
        $0.226\pm 4.0\%$ & $0.370\pm 4.8\%$ & $0.0717\pm 3.6\%$ & 
        $0.228\pm 6.1\%$ & $0.365\pm 5.0\%$ \\
    source flux density $F_{\mathrm{Q}}~\mathrm{[mJy]}$\footnote{Flux density without colour correction.} 
        & $0.50\pm 25\%$&$2.40\pm 4.3\%$ & $2.82\pm 5.5\%$ 
        & $8.33\pm 6.2\%$&$3.67\pm 5.0\%$ & $1.78\pm 7.0\%$ \\
    source flux density $F_{\mathrm{Q}}^{corr}~\mathrm{[mJy]}$\footnote{Colour-corrected according
        to a spectrum $F_{\lambda}\propto \lambda^{-2}B_{\lambda}(T)$ with 
        $T=198.7~{\rm K}$ (supernova) and $T=564.1~{\rm K}$ (star).} 
        & $0.43\pm 25\%$&$2.64\pm 4.3\%$ & $2.90\pm 5.5\%$ 
        & $8.43\pm 6.2\%$&$3.73\pm 5.0\%$ & $1.69\pm 7.0\%$ \\
        \hline
  \end{tabular}
  \end{minipage}
\end{table*}

The surface brightness of the background at these 
two wavelengths was estimated from the images calculated as described in 
Sect. 2.1, but without filtering and convolution. 
To derive the surface brightness $I_{\rm U}$ at the position 
of the supernova from the maps we took the average
value within two circular areas of radius of $5\arcsec$ 
offset $15\arcsec$ to the east and to the west of SN~1987A.
At both 
12 and $14.3~\mu{\rm m}$ the difference between the two values is less than $1\%$.
For the background brightnesses at the position of the 
star we took the average of a number of single 
measurements around the star; here the estimated uncertainties
in the background are $1.3\%$ ($LW10$) and $0.9\%$ ($LW3$).

The derived values for the source flux and size,
and the background brightnesses, are summarized 
in table \ref{fitparam}, together with the assumed value for 
the spectral index $k$ used in the determination of source size.
The table also gives the source flux densities
normalized to the surface brightness of the background.
The uncertainties quoted for the normalized flux densities
at 12 and $14.3~\mu{\rm m}$ are statistical uncertainties 
derived from the $\chi^2$-fit. 

The IR spectrum of the supernova remnant can be approximated
by a simple black body spectrum with
\mbox{$T=288\pm 16$ K} \mbox{($\chi^2_{\rm min}=0.84$)} or a modified black body spectrum 
\mbox{$F_{\lambda}\propto \lambda^{-2}B_{\lambda}(T)$} with $T=198.7^{+9.9}_{-7.8}~{\rm K}$ where a good
fit is achieved \mbox{($\chi^2_{\rm min}=0.02$)}. 
Assuming the same modified spectrum for the star we get a temperature of $T=564.1_{-25.1}^{+27.2}~{\rm K}$. The fit is
again very good
($\chi^2_{\rm min}=1.2\times 10^{-3}$). A simple black body for the star seems to be unlikely ($\chi^2_{\rm min}\approx 5.3$).

\subsubsection{Sizes of the IR-sources}

\label{sect-sourcesize}
\begin{figure}[htbp]
  \resizebox{\hsize}{!}{\includegraphics{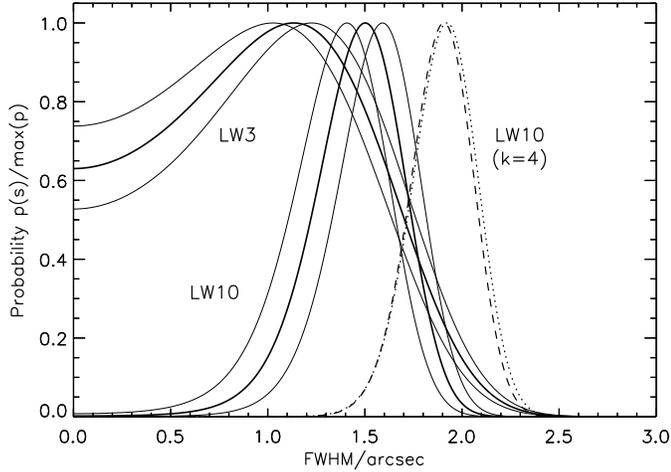}}  
  \caption{Derived probability distributions of the size $s$ of the theoretical image 
    (see Eq.~\ref{source_s}) as a function
    of the $FWHM$ of the IR source for the two filter measurements \emph{LW10} ($12~\mu{\rm m}$) and 
    \emph{LW3} ($14.3~\mu{\rm m}$) of SN~1987A. The emitting region is assumed to be a circular
    Gaussian.
    Also shown is the effect of different source spectra of form
    $F_{\lambda}\propto\lambda^{-k}$ with spectral index $k$. 
    The probability distributions shown as thick lines correspond to the most probable spectral index $k=1.08$ as 
    derived from the 
    flux densities at 12 and $14.3~\mu{\rm m}$. The distributions corresponding
    to the upper and lower limits for $k$ ($0.56$ and $1.59$, respectively, at 90\% confidence level)
    are plotted as thin solid lines.
    The probability distributions shown as dashed and dotted lines are size measurements of the
    IR emission associated with the supernova at $12~\mu{\rm m}$ using
    as $psf$s an image of a calibration star (HIC78527)  
    and a theoretical $psf$ with $k=4$, respectively.
    }
  \label{sizeprobdistr}
\end{figure}

The probability distributions for the extension of the IR-emitting regions
associated with SN~1987A were derived 
at 12 and at $14.3~\mu{\rm m}$ from a $\chi^2$-fit using 
a symmetrical Gaussian as source function
(see appendix \ref{app_sourcesize}). These are shown in Fig. 
\ref{sizeprobdistr}. 
The different probability distributions correspond to different colour
indices used in the calculation of the theoretical $psf$.
The distributions plotted with thick solid lines have been derived for the most 
probable colour of the source, a power law with $k=1.08$. 
As can be seen in the figure, the
probability distributions are only slightly affected by the uncertainty in the spectral index $k$
for the MIR emission associated with the supernova. The probability distributions corresponding
to the upper and lower limits for $k$ ($0.56$ and $1.59$, respectively, at 90\% confidence level)
are plotted as thin solid lines.
The dashed and dotted curves peaking near $1\farcs 9$ are probability distributions for the
measurements at $12~\mu{\rm m}$ using as $psf$ the image of a calibration star and the theoretical
$psf$ for a spectral index $k=4$, respectively.  Since the spectrum of the star should be
a power law with $k=4$, the similarity of these two curves demonstrates that the effect of interchanging
the theoretical $psf$ with the measured $psf$ is small.
The larger apparent source sizes
calculated from the $psf$s with $k=4$ 
arise from the decrease in $FWHM$ of the $psf$ with increasing $k$ compared with the true value $k=1.08\pm 0.32$
for the MIR emission associated with SN~1987A.

For the $LW10$ observation we obtained 
a $FWHM$ of $1.50_{-0.24}^{+0.21}$ arcsec 
for the extension of the source. This 
size is in agreement with the extensions of the projected inner ring, 
seen in the HST picture. The distribution we derived from the $LW3$ observation is broader, and
also has a higher probability for small extensions.
This might be due to the strong glitches causing long time distortions of the detector response 
of the observation (see also Sect. \ref{sourceangle}). From this observation we were only able to
obtain an upper limit for the source size of $2\farcs 3$ at $3\sigma$ level.
Table \ref{size_cl} gives
the maximum and minimum sizes of the IR-emitting region for 
certain confidence levels derived from the probability distributions for $k=1.08$.
As shown in appendix~\ref{verification} the theoretical $psf$ for the $LW10$ filter 
is a sufficiently accurate representation of the
real $psf$ to resolve sources with a $FWHM$ greater than $\sim 0\farcs 6$.
If we take the measurement at $12~\mu{\rm m}$ the source is extended with a certainty
of more than 99\%.

\begin{table}[phtb]
  \caption[t]{
    \label{size_cl}
        Sizes of the IR source associated with SN~1987A}
  \begin{tabular}{l|ccccc}
        \hline
        \hline
        conf. level & 68 \% & 90 \% & 95 \% & 99 \% & 99.9 \% \\
        \hline
        \hline
        & \multicolumn{5}{c}{Maximum size of the source ($FWHM$)} \\
        \hline
        \emph{LW3}  & $1\farcs 46$ & $1\farcs 79$ & $1\farcs 92$ & $2\farcs 15$ & $2\farcs 40$ \\
        \emph{LW10} & $1\farcs 60$ & $1\farcs 76$ & $1\farcs 83$ & $1\farcs 95$ & $2\farcs 08$ \\
        \hline
        & \multicolumn{5}{c}{Minimum size of the source ($FWHM$)} \\
        \hline
        \emph{LW3}  & $1\farcs 03$ & $0\farcs 62$ & $0\farcs 45$ & $0\farcs 21$ & $0\farcs 07$ \\
        \emph{LW10} & $1\farcs 40$ & $1\farcs 19$ & $1\farcs 09$ & $0\farcs 87$ & $0\farcs 56$\\
        \hline
  \end{tabular}
\end{table}



Some evidence was found for extended emission associated with the
star at $12~\mu{\rm m}$ (Table \ref{fitparam}).
This may in part be due to the strong
edge position and the course sampling of the star. But it cannot be excluded from the IR observations
that the star itself is extended.

\subsection{Fits with an elliptical Gauss function as IR-source}

\label{sourceangle}
\begin{figure*}
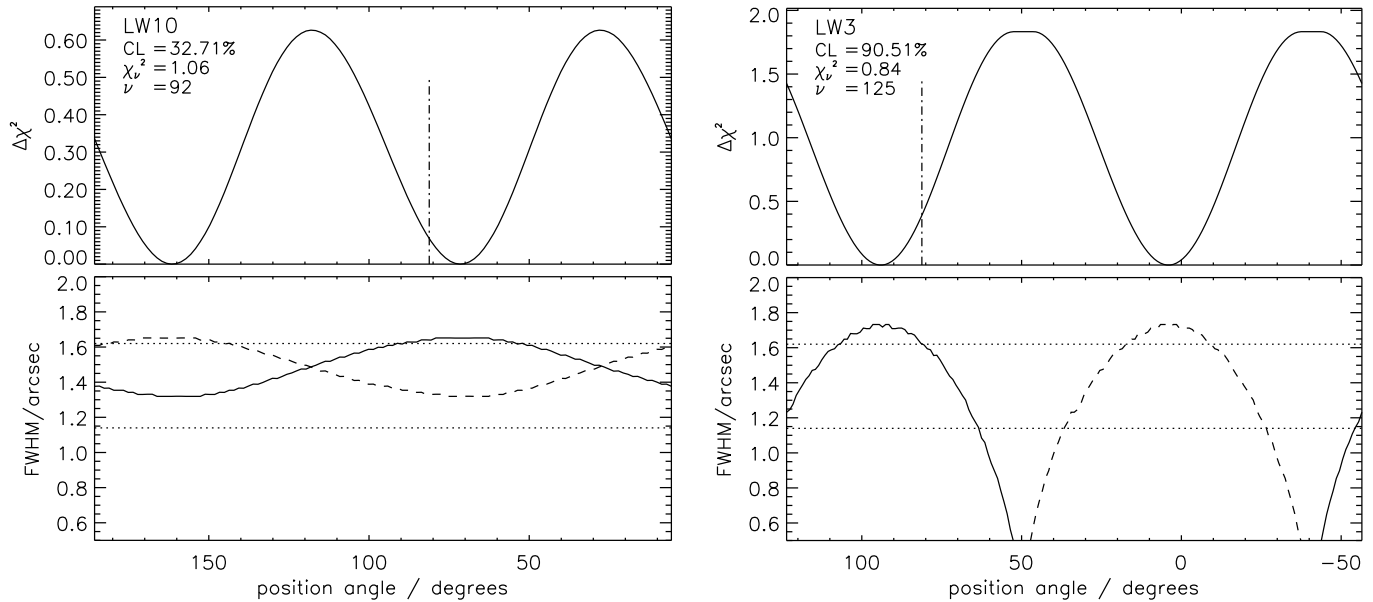

  \begin{minipage}[t]{0.49\hsize}
  \resizebox{\hsize}{!}{\includegraphics{ms1827f6.epsi}}
  \end{minipage}
  \hfill
  \begin{minipage}[t]{0.48\hsize}
  \resizebox{\hsize}{!}{\includegraphics{ms1827f7.epsi}}  
     \end{minipage}
  \caption{
    \label{chi-angle_var}
    The top panel shows the variation of $\Delta\chi^2$ with the position angle of a two 
    dimensional Gaussian
    as source function of the IR emission at $12~\mu{\rm m}$ (\emph{LW10}) and $14.3~\mu{\rm m}$ (\emph{LW3}).
    Below the $FWHM$ of the two directions of the two dimensional Gaussian (solid and dashed line) are
    shown as function of the position angle. The extensions are compared with the lengths of the two axes of the 
    elliptical projection (dotted lines)
    of the thick inner ring of SN~1987A, seen with the WFPC2 of the HST (Burrows et al., \cite{Burrows95}).
    The position angle of the major axis of the ring projection ($81.2^{\circ}$) is shown in the top panels
    as the vertical dashed-dotted lines. The given confidence levels
    belong to the minimum $\Delta\chi^2$.
    }
\end{figure*}

In addition to the extension we also examined the orientation angle of the IR-source.
Herefore we described the source through a two dimensional elliptical Gauss function with 
variances $\sigma_{\tilde x}$ and $\sigma_{\tilde y}$ and a rotation angle
$\vartheta$, which was taken to be the angle between the negative $Z$-axis in the satellite coordinate system
and the $\tilde y$-axis of the rotated Gaussian, measured counter clockwise. 
$\vec a$ is then 
$[F_{\rm Q}, \vec x_0, \sigma_{\tilde x},\sigma_{\tilde y},\vartheta,I_U]$ and the model image given by:
\begin{eqnarray}
  B(\vec x,\vec a) = I_U+ \int {\rm d}\vec x'\bigg\{
  J_{\rm F}(\vec x - \vec x') 
    \frac{F_{\rm Q}}{2 \pi \sigma_{\rm \tilde x}\sigma_{\rm \tilde y}} \quad\nonumber \\
  \times \exp\left({{-\frac{\tilde x(\vartheta,\vec x',\vec{x_0})^2}{2\,\sigma_{\rm \tilde x}^2}-\frac{\tilde y(\vartheta,\vec x',\vec{x_0})^2}{2\,\sigma_{\rm \tilde y}^2}}}\right)\bigg\},&&
\end{eqnarray}
where $\vec{ \tilde x}(\vartheta,\vec x,\vec x_0)=(\tilde x(\vartheta,\vec x,\vec x_0),\tilde y(\vartheta,\vec x, \vec x_0))$ are the coordinates of the rotated Gaussian:
\begin{equation}
  \vec{\tilde x}(\vartheta, \vec x,\vec x_0) = \left(\begin{array}{r r}
    \cos(\vartheta) & \sin(\vartheta)\\ -\sin(\vartheta) & 
    \cos(\vartheta)\end{array}\right)(\vec x- \vec x_0).
\end{equation}
As before we used theoretical
$psf$s for the two filters \emph{LW10} and \emph{LW3} corresponding to $k=1.08$.
For each rotation angle $\vartheta$ we derived best fit variances through the minimization of $\chi^2(\vec a)$ (Eq. \ref{chisqr_equation}), 
leaving
$F_{\rm Q}$ and $\vec x_0$ as free parameters and $I_U$ fixed. 
The position angle of the
elliptical Gaussian is taken to be the angle of the $\tilde y$-axis 
to the northern direction, measured again counter clockwise.
The variation
of the minimum $\Delta\chi^2$ with the position angle for the two filters
is shown in Fig. \ref{chi-angle_var}, together with the corresponding extensions given as $FWHM$.
The extensions and the position angle are compared with the orientation and the sizes of the 
minor and the major axis of the projected inner ring of SN~1987A, seen with the WFPC2 of the HST
(Burrows et al., \cite{Burrows95}). The source sizes of the best fit are
shown in Fig. \ref{IR_sourcesize} overlaid on an WFPC2-image, where we have centred the 
ellipses representing the fitted Gaussian source model at the position of the SN~1987A.

The variation in $\Delta \chi^2$ is low in the case of the measurement with \emph{LW10}. But 
the coincidence of the derived parameters of the extensions and the rotation angle of the
two dimensional Gauss function with the parameters of the inner ring are remarkable. 
The derived region of the IR source at $14.3~\mu{\rm m}$ on the other hand
seems to have an extension with the dimension of the inner ring only in one 
direction. In the other direction, the shown extension is the minimum we allowed to
avoid inaccuracies in the modeling. We think that this strong asymmetry is artificial and an
effect due to strong glitches at the beginning of the \emph{LW3} observation.

\subsection{Position of the individual sources}

\label{coordinatesection}

\begin{figure*}[htbp]
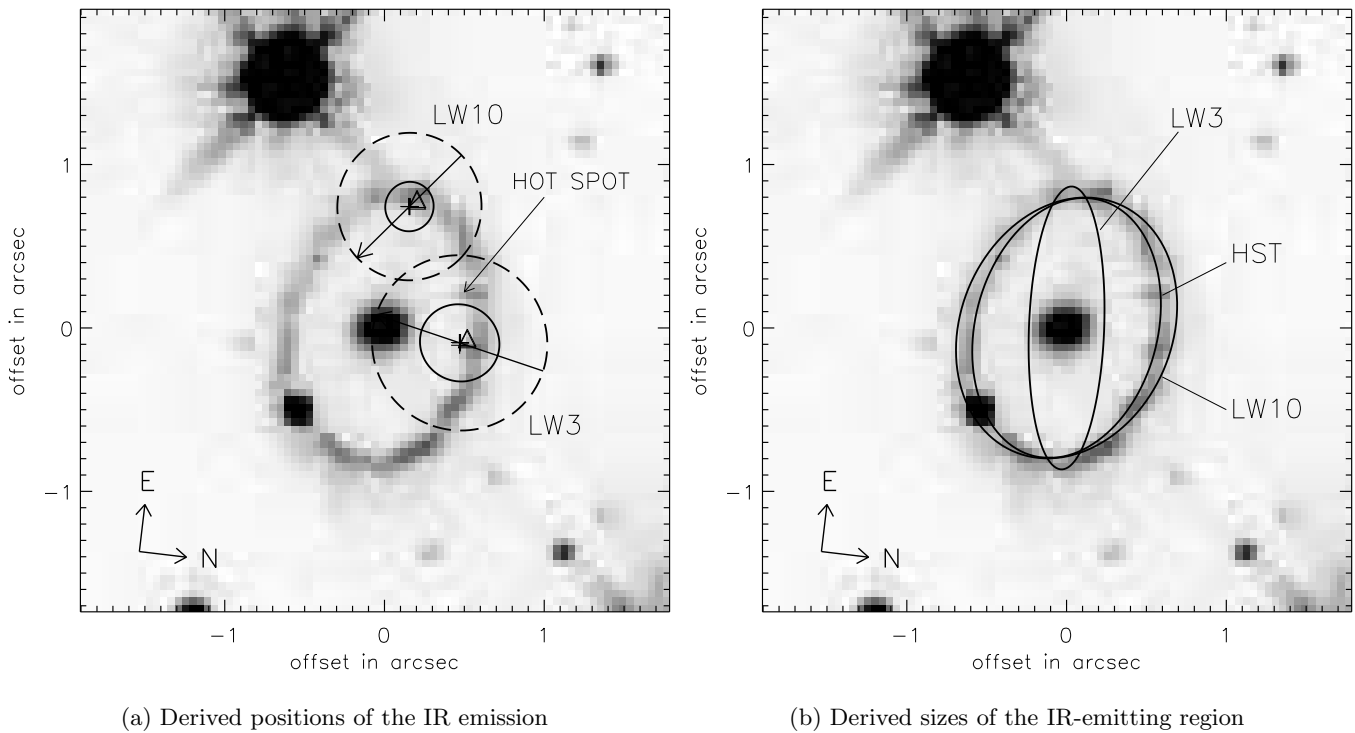

 \begin{minipage}[t]{\hsize}
  \subfigure[Derived positions of the IR emission]{
    \label{IR_positions}
    \resizebox{0.49\hsize}{!}{\includegraphics{ms1827f8.epsi}}}
  \hfill
  \subfigure[Derived sizes of the IR-emitting region]{
  \label{IR_sourcesize}
  \resizebox{0.49\hsize}{!}{\includegraphics{ms1827f9.epsi}}}
  \caption{Derived absolute coordinates and sizes of the IR-emitting region 
    of the measurements at $12~\mu{\rm m}$ (\emph{LW10})
    and $14.3~\mu{\rm m}$ (\emph{LW3}), overlaid on an HST image of SN~1987A, made in V-band with 
    the WFPC2 on 6. February 1996 (showing the thick inner ring and emission
    from the fast expanding ejecta inside). The
    point source to the east is an unrelated star (after convention \mbox{Star 3}, 
    e.g. Walborn et al., \cite{Walborn93}). The bright spot on the ring to the south-west is
    a star in the line of sight.}
    (a) The coordinates of the infrared emission, derived on the basis of NTT and HST data, are shown
    as crosses and triangles, respectively.
    The almost circular ellipses, drawn with a solid and a broken line, respectively, show the statistical errors 
    ($\sim 0\farcs 15$ and \mbox{$\sim 0\farcs 24$}, respectively at 12 and $14.3~\mu{\rm m}$) and
    the total inaccuracy of the derived positions of the IR-emission.
    The systematic error is not well known, because of the edge position of the star, and is simply
    assumed to be given by $0\farcs 42$, which is the combined uncertainty of the pixel coordinates
    in the inner ISOCAM field after correcting for distortion (Aussel, \cite{Aussel98}).
    The arrows indicate the orientations of the $Z$-axis of the ISO--satellite.
    Also shown is the position of the so-called \emph{Hot Spot}, 
    where at least since 1995 the emission in optical wavelengths 
    at the inner boundary of the inner ring started to brighten (Lawrence et al., \cite{Lawrence00}).\\
    (b) The derived sizes of the emitting regions for the two IR measurements with the \emph{LW10} and \emph{LW3}-filter
    are drawn as ellipses centred on the position of SN~1987A. Each of the two axes of the ellipses correspond to the
    $FWHM$ of the two dimensional Gaussian, determined for the IR-source. The orientation and size of
    the derived size of the IR-emission at $12~\mu{\rm m}$ agrees very well with the projected
    ellipse of the inner ring in the HST-picture, which is outlined, using the orientation and sizes
    given by Burrows et al. (\cite{Burrows95}).
  \end{minipage}
\end{figure*}

As an independent property of the IR emission from the remnant of SN~1987A 
we derived the absolute coordinates relative to the position of the IR-emitting star
visible in the two mosaics at 12 and even at $14.3~\mu{\rm m}$ (Fig. \ref{ISOCAM-fig}).
The position of the IR-emitting star itself was not known so that we first had to determine
the absolute coordinates of this star (see appendix \ref{starposition}). 

We derived the following offset positions of the IR emission from the position of the supernova, 
where we have taken the coordinates given by Reynolds et al. (\cite{Reynolds95}):
\begin{equation}
  \begin{array}{lcl}
    \Delta \alpha &=& -0\farcs 04\pm 0\farcs 24_{\rm stat.}\pm 0\farcs 42_{\rm sys.},\\ 
    \Delta\delta &=& +0\farcs 48\pm 0\farcs 24_{\rm stat.} \pm 0\farcs 42_{\rm sys.}  
  \end{array}
  \quad \mathrm{({LW3})}
\end{equation}
and 
\begin{equation}
  \begin{array}{lcr}
    \Delta \alpha &=& +0\farcs 77\pm 0\farcs 15_{\rm stat.} \pm 0\farcs 42_{\rm sys.}, \\ 
    \Delta\delta &=& +0\farcs 09\pm 0\farcs 15_{\rm stat.} \pm 0\farcs 42_{\rm sys.}.
  \end{array}
  \quad \mathrm{({LW10})}.
\end{equation}
The systematic uncertainty is not very well known.
    It is simply assumed
    that the systematic uncertainty in individual directions is given by the accuracy
    of the pixel positions in the inner ISOCAM-field
    after distortion correction, which should be accurate to within a tenth of a
    pixel (Aussel, \cite{Aussel98}). The combined
    uncertainty of the coordinate pair is therefore taken to be $\sim 0\farcs 42$. 
Due to the strong edge position of the star in particular in the \emph{LW10}-field and the fact,
that at one raster point the dead line of the \emph{LW} camera of the \emph{LW10} observation 
was lying at the sky position of the star, the systematic uncertainty of the offset position 
at $12~\mu{\rm m}$ may be larger. 

The absolute coordinates of the IR-emission from the remnant of SN~1987A at 12 and $14.3~\mu{\rm m}$
with the corresponding uncertainties are shown in Fig. \ref{IR_positions} in relation
to the WFPC2-image. Fig. \ref{IR_sourcesize} shows a visualization of the fits to the 
size of the emitting region. 
As seen in Fig. \ref{IR_positions} the
IR emission measured with ISOCAM should be from of the remnant of SN~1987A.
Taking the joint probability 
distribution of both measurements, the most probable location for the IR counterpart
is to the north east 
of the supernova, which is also the position of the maxima in the 
X-ray and radio surface brightness distributions  (Burrows et al., \cite{Burrows00}).
But it is also possible that the centre of the IR emission lies 
at the position of the supernova so that the outer boundary of the emitting region could indeed
coincide with the inner ring as seen in Fig. \ref{IR_sourcesize}.
Therefore there is no firm evidence of asymmetry in the MIR emission comparable with that seen in radio
and X-ray wavelengths.

\section{Discussion}

Very probably the detected MIR flux densities are dominated by emission from dust grains.

The contribution of synchrotron radiation can be estimated 
by extrapolating the radio spectrum given by Gaensler et al. (\cite{Gaensler97}) to
$10~\mu{\rm m}$. Taking into account the linear increase of radio emission with time,
this yields the value $\sim 5.1\times 10^{-3}~{\rm mJy}$, 
two orders of magnitude less than the flux densities measured by ISOCAM.

We can also exclude free-free emission from the 
debris or the thick inner ring as a substantial contributor to the observed
MIR flux densities. We have estimated the MIR free-free continuum from optical recombination
line measurements using published emissivities and recombination coefficients
for case B of Martin (\cite{Martin88}).

For the free-free emission from the ejected material 
we adopted the ${\rm H}_{\beta}$ flux 2873 days after outburst
(Chugai et al., \cite{Chugai97}) assuming a temperature of \mbox{$10^3$ K}, which is consistent with
the measured balmer decrement of 3.8. 
The predicted flux density at $10~\mu{\rm m}$ is roughly  
three orders of magnitudes lower ($\sim 3.2\times 10^{-3}~{\rm mJy}$). 
Even in the case of a higher temperature of $10^4$ K the free-free
emission from the supernova debris in the IR is negligible ($\sim 0.027~{\rm mJy}$). 

The predicted free-free emission from
the thick inner ring is brighter, but still insufficient to account for the ISOCAM detections.
To estimate its contribution to the IR the temperature of the whole inner ring was taken to be 
$10^4$ K as was derived from analysis of emission spectra, 
made with the HST at different epochs from a bright part of the inner ring
(Lundqvist \& Sonneborn., \cite{Lundqvist97}). 
We used the flux of the inner ring in R-band, measured with the WFPC2 of the HST 3268 days after outburst 
(Soderberg et al., \cite{Soderberg99}). Further, we assumed that 19\% of the flux is due to the ${\rm H}_{\alpha}$-line
and the rest to the neighbouring NII lines at 6548 and 6583 \AA{}, 
which is roughly consistent with the line ratios 
in the line spectrum of a luminous part of the ring presented by Panagia et al. (\cite{Panagia96}). 
For the flux density of the free-free emission at $10~\mu{\rm m}$ from the inner thick ring
we derived an upper limit of $\sim 0.053~{\rm mJy}$.

The contribution of the synchrotron radiation, the free-free emission of the ring and the debris to
the IR is shown in
Fig. \ref{SED}, which summarizes the spectral energy distribution of the SN~1987A 
at late times. 

No sensitive observations of the MIR cooling lines at late epochs are available. However,
according to the theoretical predictions for the evolution of
the emission from various fine structure lines of the supernova debris 
(Kozma \& Fransson, \cite{Kozma98b}) their contribution should be negligible. 
We conclude that 
the measured IR emission is most probably thermal emission from dust. 
This emission could either be due to grains in the shocked gas in the circumstellar environment or
originate in condensates heated mainly by the energy of the radioactive decay of
the nucleosynthesis products.


\begin{figure}[htbp]
  \resizebox{\hsize}{!}{\rotatebox{90}{
      \includegraphics{ms1827f10.epsi}}}
  \caption{
    \label{SED}
    Spectral energy distribution of the SN~1987A at late times. 
    The measured ISOCAM fluxes are shown as open diamonds. They are colour-corrected due to the
    modified black body spectrum $F_{\lambda}\propto B_{\lambda}(T)\lambda^{-1}$
    with a temperature of $236.1~{\rm K}$ 
    taken as approximation for the circumstellar dust emission (solid line). 
    Upper limits in broad band at 25, 60, 100, 170 and $200~\mu{\rm m}$ (short horizontal lines) 
    are measurements with ISOPHOT made at the same epoch as
    the ISOCAM observations (Fischera, \cite{Fischera00}).
    The open circles
    show measured fluxes at 1.3 mm 1289 and 1645 days after outburst (Biermann et al., \cite{Biermann92}). 
    The L$\alpha$ fluxes 3715 and 3734 days after outburst
    are taken from Sonneborn et al. (\cite{Sonneborn98}). 
    Upper limits, derived from measurements with the SWS-spectrometer aboard ISO and shown as filled circles, 
    are taken from Lundqvist et al. (\cite{Lundqvist99}). 
    Optical fluxes from the thick inner ring (triangles) and from the debris (squares), 
    derived from observations with the WFPC2 of the HST, 
    are taken from Soderberg et al. (\cite{Soderberg99}). 
    The synchrotron spectrum is an extrapolation from radio measurements, published  by Gaensler et al.
    (\cite{Gaensler97}). For the Bremsstrahlung in X-ray (solid line) we assumed a temperature of 
    $2\times 10^7$ K and scaled the emission measure 
    $\sim 2500$ days after outburst ($EM=(1.4\pm0.4)\times 10^{57}~{\rm cm^{-3}}$, 
    Hasinger et al. (\cite{Hasinger96})) with $EM\propto t^{2.06}$
    to the time of the ISOCAM observations.
    The free-free emission of the thick inner ring and the debris are derived from
    line fluxes, given by Chugai et al. (\cite{Chugai97}), and the flux in R-band, given by 
    Soderberg et al. (\cite{Soderberg99}). The lines in IR are FeII cooling lines from the
    ejecta at late times derived from theoretical studies for $10^{-4}M_{\sun}$ of $^{44}{\rm Ti}$ 
    (taken from Lundqvist et al., \cite{Lundqvist99}). In particular the line at $26~\mu{\rm m}$,
    which is thought to be the strongest cooling line at late epochs, is of basic interest 
    as a tracer of the mass of $^{44}{\rm Ti}$. Using this line
    the $3\sigma$ upper limit of 51~mJy of the ISOPHOT broad band measurement
    at $25~\mu{\rm m}$ corresponds to a $3\sigma$ upper limit of $\sim 1.5\times 10^{-4}~M_{\sun}$ for the
    mass of $^{44}{\rm Ti}$
    (Fischera, \cite{Fischera00}).
    The emission from condensates is expected to be mainly in the FIR (see text).} 
\end{figure}

To derive dust masses knowledge of the optical properties of the grains is needed.
The grain emissivity can depend quite strongly on grain composition and in particular in the ISOCAM range
on wavelength (see e.g. Draine \& Lee, \cite{DraineLee84}). Nevertheless,
for a first interpretation the dust spectrum may be approximated by a modified blackbody function
of the form already used for the colour correction in Sect.~\ref{basicfitresults}.
The flux density of $N(a)$ dust grains with radius $a$ at 
a distance $D$ radiating at a certain temperature $T$ is then given by
\begin{equation}
  F_{\lambda}= N(a)\frac{a^2}{D^2}\,\pi \,B_{\lambda}(T) \,Q_{\lambda}(a),
\end{equation}
where the emission coefficient is assumed to be $Q_\lambda(a)=a\,\xi\,\lambda^{-\beta}$
with constant factors $\xi$ and $\beta$.
The total luminosity of such a spectrum is given by:
\begin{eqnarray}
  L &=& N(a)\,4\pi a^2\,\sigma T^4 \left<Q(a,T)\right> \\&=& N(a)\,4\pi a^2\,\sigma T^4 \,\frac{\Gamma(4+\beta)\zeta(4+\beta)}{\Gamma(4)\zeta(4)}
    \left(\frac{k T}{h c}\right)^\beta a \,\xi,
\end{eqnarray}
where $k$, $h$, $c$ and $\sigma$ are the Boltzmann constant, the Planck constant, the velocity of 
light and the Stefan--Boltzmann constant. $\left<Q(T,a)\right>$ is the Planck--averaged emission coefficient
and $\Gamma(x)$ and $\zeta(x)$ are the gamma- and the zeta-functions.
Whereas the emission behaviour of spherical small grains at long wavelengths can be described with $\beta=2$
at shorter wavelengths a better approximation is given by $\beta=1$. The fit of a spectrum with $\beta=1$
including a colour correction to the flux densities measured with ISOCAM is shown in Fig.~\ref{SED}. 
The corresponding temperature is $236.0\pm 11.5$~K and the luminosity of the best fit is
$L=2.47\times 10^{28}~{\rm W}$. 
As seen in the figure the approximation gives a good
explanation of the observed colours ($\chi^2=0.14$). 

The luminosity can be compared with 
the rate of deposition of energy from the decay of radioactive decay products.
Assuming the ejecta after 11 years to be optically thin to gamma rays 
and taking an initial mass of $10^{-4}~M_{\sun}$ of $^{44}{\rm Ti}$ which is mainly responsible
for the heating at late epochs
then the luminosity emitted by dust grains in the MIR is at least $\sim 20\%$ of the 
deposited energy rate $L\approx 1.23\times 10^{29}~{\rm W}$ (Fischera, \cite{Fischera00}).

The dust mass was estimated from the equation
\begin{equation}
  M_{\rm dust} = \frac{L\rho}{3\,\sigma T^4 \,(\left<Q(T,a)\right>/a)}
\end{equation}
where $\rho$ is the grain density. We used Planck--averaged emission coefficients
at 236~K of silicate and graphite grains with a radius $a=0.01~\mu{\rm m}$ 
adopted from published values from Laor \& Draine (\cite{LaorDraine93}) which are
$\left<Q_{\rm Si}\right>/a[\mu{\rm m}]\approx 0.42$ and 
$\left<Q_{\rm Gra}\right>/a[\mu{\rm m}]\approx 0.080$. The grain densities are taken
to be $2.3~{\rm g/cm^3}$ for graphite and $3.2~{\rm g/cm^3}$ for silicate grains.
The masses of silicate and graphite grains are found to be 
$\sim 1.7\times 10^{-7}~M_{\sun}$ and $\sim 6.8\times 10^{-7}~M_{\sun}$ independent of grain size.

These values indicate a dust mass which is much
lower than the minimum dust mass of $\sim 10^{-4}~M_{\sun}$ derived for the newly formed
dust in the metal rich part of supernova ejecta (Wooden, \cite{Wooden97}). It is probable that the bulk
of the condensates are currently emitting in the FIR/submm and were too cold to be visible
to ISOCAM.
A temperature of at least $\sim 200$ K is needed to explain
the colours of the ISOCAM measurement. By contrast,
even as soon as $\sim 1300$ days after outburst, the inferred temperature of condensates was only
$\sim 140~{\rm K}$ (Bouchet et al. (\cite{Bouchet91}), Biermann et al., (\cite{Biermann92})). The condensate 
temperature should,
due to the expansion of the ejecta and weaker heating, have been even lower during the
ISOCAM observation $\sim 4000$ days after outburst. The bulk of the condensates should therefore
mainly emit in the FIR. Energetically, it is feasible that 
the observed MIR emission could nevertheless be attributed to a small fraction of
condensates residing in clumpy regions rich in
radioactive nucleosynthesis products where the local heating would be stronger. 
However, detailed modeling of this case shows that the MIR spectral shape would only be reproduced 
for a small range of filling factors, composition and sizes 
(Fischera, \cite{Fischera00}).

The fact that the MIR-emission was resolved by ISOCAM strongly suggests that the bulk of the 
detected emission does not arise from condensates. 
The condensates are thought to be restricted to the central metal 
rich core of the ejecta (see e.g. Wooden, \cite{Wooden97}), 
that is expanding with a velocity of $2000~{\rm km/s}$ 
(Kozma \& Fransson, \cite{Kozma98}). 
At the time of the ISOCAM observations $\sim 4000$ days after outburst the condensates did not
cover a region larger than $\sim 0\farcs 18$. 


The derived size of $1.50_{-0.24}^{+0.21}$ arcsec and the orientation of the emitting region, with
an extension consistent with the elliptical projection of the
thick inner ring (Sect. \ref{sourceangle}), both point to a
predominantly circumstellar origin for the dust emission.
The MIR extent is furthermore consistent with the extensions
of the radio (Gaensler, \cite{Gaensler97}) and X-ray (Burrows, \cite{Burrows00}) 
emitting regions. 
The presence of dust in the circumstellar environment has also been concluded from observations
of scattered light correlated with the thick inner ring and the associated nebula
(Wampler et al., \cite{Wampler90}; Crotts et al., \cite{Crotts95}).

Due to the fact that both the radio and the X-ray emission are thought to be associated with the blast wave as it propagated 
into an HII-region of moderate density interior of the thick inner ring (Chevalier \& Dwarkadas, \cite{Chevalier95}), 
it is likely that the emission in the MIR arises from circumstellar dust heated by shocked gas.
The observed dust temperature of $\sim 200$~K is consistent with the predictions for collisionally heated grains
in equilibrium with a surrounding hot gas with a density of several $100~{\rm cm^{-3}}$ (Fig.~7 in Dwek, \cite{Dwek87}).
Gas density of this order is indeed inferred from the observed X-ray emission in the shocked circumstellar medium of
SN~1987A (Borkowski et al., \cite{Borkowski97}). It is therefore to be expected that
the MIR emission should have a detailed morphology similar to the ring-like structure as seen in X-rays 
(Burrows, \cite{Burrows00}) with the centre close to the position of the supernova. 
The derived luminosity in MIR is higher than in X-rays (Hasinger et al., \cite{Hasinger96})
by a factor of $\sim 10$ as seen also in Fig.~\ref{SED}. 
This is in agreement with values typically observed in galactic SNRs (Fig. 6 in Dwek \cite{Dwek88}).

The presence of dust in the circumstellar environment is consistent with a red supergiant
phase in the time evolution of the supernova progenitor (Woosley, \cite{Woosley88}). In particular,
our interpretation of the IR emission is in agreement with the hypothesis that the HII-region
is wind material from this red supergiant phase as suggested by Chevalier \& Dwarkadas (\cite{Chevalier95}).



\section{Summary}
We have presented measurements made with ISOCAM at $6.75$, $12$ and at 
$14.3~\mu{\rm m}$  $\sim 11$ years after outburst of SN~1987A. To characterize
the IR source we derived the absolute coordinates and analysed 
the extension of the emitting region. 

\begin{enumerate}
  \item The flux densities of the MIR source associated with the supernova 
    at 6.75, 12 and at $14.3~\mu{\rm m}$ are $0.496\pm25.1\%$ mJy, $2.41\pm4.3\%$ mJy 
    and $2.82\pm 5.5\%$ mJy (respectively 3855.2, 3998.7 and 3937.7 days after outburst) (Sect. \ref{sectionparameter}).
  \item The offsets of the IR emission of the remnant of SN~1987A
    at 12 (\emph{LW10}) and at $14.3~\mu{\rm m}$ (\emph{LW3}) from the position of the supernova
    ($\mathrm{\alpha=5h35m27.968s,\,\delta=-69^{\circ}16'11\farcs 09}$ (J2000), 
    Reynolds et al., \cite{Reynolds95}),
    derived relative to an IR-emitting star in the ISOCAM fields (see Sect. \ref{coordinatesection})
    are found to be:
    \begin{equation}
      \begin{array}{lcl}
        \Delta\alpha & = & -0\farcs 04\pm0\farcs 15_{\rm stat.} \pm 0\farcs 42_{\rm sys.}, \\
        \Delta\delta & = & +0\farcs 48\pm0\farcs 15_{\rm stat.} \pm 0\farcs 42_{\rm sys.}
      \end{array}
        \quad\mbox{(LW3)}
    \end{equation}
    and
    \begin{equation}
      \begin{array}{lcl}
        \Delta\alpha & = & +0\farcs 77\pm0\farcs 24_{\rm stat.}\pm 0\farcs 42_{\rm sys.}, \\
        \Delta\delta & = & +0\farcs 09\pm0\farcs 24_{\rm stat.}\pm 0\farcs 42_{\rm sys.}.
      \end{array}
      \quad \mbox{(LW10)}
    \end{equation}
    Due to the edge position of the star the
    uncertainty of the measurement at $12~\mu{\rm m}$ may be larger. 
    On the basis of the 
    absolute coordinates taken in isolation the MIR emission could be associated either with
    the supernova or the circumstellar interaction region seen at radio and X-ray wavelengths.
  \item The absolute coordinates of the IR-emitting star derived using 
    observations of the SN~1987A made with HST and NTT are (with a 1 $\sigma$ uncertainty of $\sim 0\farcs 09$):
    \begin{displaymath}
      \mathrm{\alpha=5h35m18.418s,\,\delta=-69^{\circ}16'30\farcs 65}.
    \end{displaymath}
  \item From the observation at $12~\mu{\rm m}$ the size of the IR-emitting region, assuming a 
    spherical symmetric Gauss function for the source, is found 
    to be $1.50_{-0.24}^{+0.21}$ arcsec ($FWHM$), consistent with the diameter of $1.66\pm 0.03$ arcsec
    found by Panagia et al. (\cite{Panagia91}) for the thick inner ring.
    Further analysis of these IR data with an elliptical Gauss function also shows a remarkably 
    good agreement of the IR-emitting region with the orientation and the extension of the projected ring
    even though the result is not very significant.
    The measured extension is a clear indication of a predominantly circumstellar origin for the IR emission. 
    In particular the emission from condensates is not prominent, because the condensates would be 
    restricted after 4000 days to a region with a diameter of $\sim 0\farcs 18$.
\end{enumerate}

\begin{acknowledgements}
The work was supported by 
Deutsches Zentrum f\"ur Luft- und Raumfahrt e.V. (DLR) through the projects 
`50 OR 9702' and `50 OR 99140'. 
The ISOCAM data presented in this paper were analysed using `CIA', a joint development by the ESA Astrophysics Division and the ISOCAM Consortium. The ISOCAM Consortium is led by the
ISOCAM PI, C. Cesarsky. The paper is
based on observations made with {\em ESO} Telescopes at the La Silla or Paranal Observatories under program
     ID 000.0-0000 (observation day: 10/01/95, observer: S. Benetti). It is further based
on observations made with the {\em NASA/ESA} Hubble Space Telescope, 
obtained from the data archive at the Space
Telescope Institute. STScI is operated by the association of Universities 
for Research in Astronomy, Inc. under the {\em NASA}
contract  NAS 5-26555. 
We thank Ren\'e Gastaud, Stephan Ott, Leo Metcalf and Ralf Siebenmorgen for discussions and information
relevant to the analysis of the ISOCAM data.
\end{acknowledgements}

\appendix

\section{Theoretical Resolution function}

\label{psfsection}
To estimate the flux, size and coordinates of the IR-source we made use of a theoretical
point spread function ($psf$) for the ISO-satellite kindly made available to us by Dr. Ralph Siebenmorgen.
Images of point sources observed 
in narrow band \emph{LW}-filters of ISOCAM have been shown to be (except for \emph{LW1}) in good agreement with this
function, which we will refer to as the monochromatic $psf$ (Okumura, \cite{Okumura98}). 
For the broad-band filters the $psf$ 
depends however on the source spectrum.
Therefore to achieve a high accuracy
in the determination of the source extension
we used a theoretical broad-band $psf$,
given by:
\begin{equation}
  J_{\mathrm{F}}({\vec x})=
  \frac{\int d\lambda\,\lambda\,R(\lambda)\,J(\vec x,\lambda)}
  {\int d\lambda\,\lambda\,R(\lambda)\,F_{\lambda}}.
\end{equation}
Here $R(\lambda)$ is the spectral transmission of an ISOCAM filter, taken from the CIA program and
$F_{\lambda}$ the flux density (assumed
to have a power law $F_{\lambda}\propto\lambda^{-k}$). $J(\vec x,\lambda)$ is given by
\begin{equation}
  J(\vec x,\lambda)=\frac{A\,F_{\lambda}}{\lambda^2}\,psf(\vec x,\lambda),
\end{equation}
where $A$ is the effective area of the telescope and $psf$ the monochromatic point spread function scaled to
unity at \mbox{$\vec{x}=0$}. $J({\vec x,\lambda})$ was derived on a grid of $2048 \times 2048$ grid points sampled
at $1/3\arcsec \times 1/3\arcsec$ using a FFT technique.
We used values of 60~cm, 17.4~cm and 2~cm for the diameter of the primary mirror, 
the borehole and the thickness of the braces of the tripod, respectively.

Finally, we convolved the obtained function $J_{\rm F}(\vec x)$ with the used pixel size and
the probability distribution of the actual pointing position of the satellite 
during the observation ({\em jitter}).

\begin{table}
\begin{minipage}[]{\hsize}
  \renewcommand{\thefootnote}{\it\alph{footnote}}
  \caption{\label{comparison_fwhm}
    $FWHM$ of the measured and the theoretical $psf$}
  \begin{tabular}{l||c|c}
     \hline
     \hline
     Filter & \emph{LW10} & \emph{LW3} \\
     \hline
     \hline
       \multicolumn{3}{c}{$FWHM$ of cuts through the maximum} \\
     \hline
     ${FWHM_{\rm meas.}[\arcsec ]}$\footnote{Values taken from Okumura, \cite{Okumura98}.}   
         & $3.981\pm 0.084$\footnote{The given uncertainties correspond
           to the difference in the two values derived in Y and Z direction.} 
         &  $4.980\pm 0.012$    \\
     ${FWHM_{\rm meas.}[\arcsec ]}$\footnote{Derived from observations of HIC78527 
     (\emph{LW10}) and 
       HIC80331 (\emph{LW3}).}  &   $3.951\pm 0.252$   & $4.980\pm 0.063$   \\
     ${FWHM_{\rm theor.}[\arcsec ]}$  &   $3.945\pm 0.026$   & $4.956\pm 0.058$   \\
     \hline
     relation meas./theor.\footnote{Due to values from Okumura.} & $1.009 \pm 0.022$  & $1.005 \pm 0.005$  \\
     \hline
        \multicolumn{3}{c}{$FWHM$ of a Gaussian approximation} \\
     \hline
     ${FWHM_{\rm meas.}[\arcsec ]}$\footnotemark[3]   & $4.071$ & 4.896\\ 
     $ {FWHM_{\rm theor.}[\arcsec ]}$ & $4.035$ & 4.866\\
    \hline
     relation meas./theor. & 1.009 & 1.006\\
     \hline
  \end{tabular}
\end{minipage}
\end{table}

To measure source extensions smaller than the
resolution of the instrument it is important to know how accurately the theoretical $psf$
represents the actual $psf$, particularly in the kernel.
To this end 
we compared the $FWHM$ of the theoretical $psf$ of the \emph{LW10} and the \emph{LW3} filters (for spectral index $k=4$) 
with the actual $FWHM$
derived from observations of point sources.
The $FWHM$ of cuts in Y and Z (in satellite coordinates) through the peak of the observed brightness
distribution was found to be in agreement to within
$1\%$ with the $FWHM$ of the theoretical $psf$. A corresponding comparison between Gaussian  fits to the observed
point source and the theoretical $psf$ also agrees within $1\%$ (Table \ref{comparison_fwhm}).
The asymmetry of the true $psf$ in the first ring visible in the images of point-like sources
(Okumura, \cite{Okumura98}) should have only a very minor effect.

An uncertainty of less than $1\%$ is smaller than the effect of varying
the spectral shape in the calculation in the $psf$. For Gaussian approximations to the
theoretical $psf$ we found that the dependence of the $FWHM$ 
on the spectral index for the \emph{LW3} and \emph{LW10} filters 
is close to linear. In a broad range from $k\approx-4$ to $k\approx4$ the $FWHM$ is approximately given by
(Fischera, \cite{Fischera00}):
\begin{equation}
  \label{FWHM}
  \begin{array}{lclll}
   {FWHM}(k) & = & 4\farcs 349-0\farcs 081\,k\quad ({LW10}),\\
   {FWHM}(k)  & = & 5\farcs 078 - 0\farcs 048\,k \quad (LW3). 
  \end{array}
\end{equation}
If the theoretical $psf$ of the calibration star ($k=4$) would also be taken for the IR emission from
the supernova remnant (which in reality has $k\approx 1$) the $FWHM$ of the $psf$ would be 
underestimated by $\sim 6\%$ and $\sim 3\%$ for the \emph{LW10} 
and \emph{LW3} filters, respectively. This can cause substantial systematic
uncertainties in the derived source sizes (Fig.~\ref{sizeprobdistr}; Sect.~\ref{verification}). 



\section{Method for determination of the probability distribution in source size}

\label{app_sourcesize}
To derive the probability distributions shown in Fig. \ref{sizeprobdistr} we measured the variation
of $\chi^2(\sigma_{\rm Q})$ as function of the variance $\sigma_{\rm Q}$ of the symmetrical Gauss function  
assumed for the source. We also left flux density ($ F_{\rm Q}$) and coordinates ($\vec x_0$) as free variables. 
Because the source is smaller than the resolution of the
telescope, we had to convolve the source function with the theoretical $psf$
described in the previous section. To transform the variation of $\chi^2$ into a probability
distribution we approximated the $psf$ itself through a two dimensional 
symmetrical Gaussian function with variance $\sigma_{\rm J}$. The value of $\sigma_{\rm J}$ was found through
a simple $\chi^2$-fit between the $psf$  and the Gaussian.
Measuring the most probable size $\sigma_{\rm Q_0}$ of the source is then equivalent to measuring the 
size of the image with the variance $s_0=\sqrt{\sigma_{{\rm Q}_0}^2+\sigma_{\rm J}^2}$. In the case
of normal distributed errors 
the probability distribution of $s$ with 
\begin{equation}
   \label{source_s}
   s = \sqrt{\sigma_{\rm Q}^2+\sigma_{\rm J}^2}
\end{equation}
is a Gaussian distribution with the variance $\sigma_s$:
\begin{equation}
  \label{gaussdistr}
  p(s)=\frac{1}{\sqrt{2\pi}\sigma_{\rm s}}\,\exp\left(-\frac{1}{2}\left(\frac{s-s_0}{\sigma_{\rm s}}\right)^2\right),
\end{equation}
so that the variation of $\Delta\chi^2(s)$ is a simple quadratic function:
\begin{equation}
  \Delta \chi^2 (s)= \chi^2(s)-\chi^2(s_0)=\frac{1}{\sigma_{\rm s}^2}(s-s_0)^2,
\end{equation}
where $\chi^2(s_0)=\chi_{\rm min}^2(s)$.
The measured variation of $\chi^2(\sigma_{\rm Q})$ can therefore be approximated by:
\begin{equation}
  \chi^2(\sigma_{\rm Q})=\frac{1}{\sigma_{\rm s}^2}
  \left(\sqrt{\sigma_{\rm Q}^2+\sigma_{\rm J}^2}-s_0\right)^2 +\chi^2(s_0).
\end{equation}
The most probable value $s_0$, its uncertainty $\sigma_{\rm s}$ and the corresponding value of $\chi^2(s_0)$
we derived again through a $\chi^2$-fit. Due to uncertainties in the model and the data the value $s_0$
can also be smaller than $\sigma_{\rm J}$.

To get the probability distribution over the physical sizes of the source we
normalized the distribution with:
\begin{equation}
   N = \int_{\sigma_{\rm J}}^{\infty}ds\,p(s)=\frac{1}{2}\left(1-erf\left(\frac{\sigma_{\rm J}-s_0}{\sqrt{2}\sigma_{\rm s}}\right)\right)
\end{equation}
using the error function:
\begin{equation}
  erf(y)=\frac{2}{\sqrt{\pi}}\int_0^y{\rm d}t\,e^{-t^2}.
\end{equation}
The confidence levels of the maximum sizes of the source are simply given by:
\begin{equation}
  \label{conflevel}
  CL(\sigma_{\rm Q}) = \frac{1}{2N}\left(erf\left(\frac{s-s_0}{\sqrt{2}\sigma_{\rm s}}\right) 
    -erf\left(\frac{\sigma_{\rm J}-s_0}{\sqrt{2}\sigma_{\rm s}}\right)\right),
\end{equation}
where $s$ is given by Eq.~(\ref{source_s}).

\subsection{Accuracy of the method}

\label{accuracy}
The accuracy of the used method to derive the extension of the emitting region can be estimated
by assuming that IR source and $psf$ are given by 
Gauss functions with variances $\sigma_{\rm Q}$
and $\sigma_J$. An uncertainty $\Delta\sigma_J$ in the variance of the $psf$
limits a measurement of the extension, 
neglecting uncertainties in the data, to sources that are larger than
$\sigma_{\rm Q}\approx \sqrt{2\Delta\sigma_J\sigma_J}$.
The uncertainties in the $FWHM$ of the theoretical $psf$ as stated in table \ref{comparison_fwhm}
gives a resolution limit of $\sim 0\farcs 6$ ($FWHM$) for both filters as will be verified in the next section.


\subsection{Verification of the method using calibration stars}

\label{verification}
To verify the method for measuring the angular size of the MIR emission associated with SN~1987A we did two independent tests.
Firstly, we made a control experiment by applying the same method to observations of point-like source
calibration stars (Sect.~\ref{test1}). Secondly, we repeated the procedure on the supernova data, but using the image
of a calibration star as the $psf$ in place of the theoretical $psf$ (see Sect. \ref{test2}).
The results both show that the model of the theoretical $psf$ allows a measurement of the size of a 
Gaussian source with a precision which corresponds to an inaccuracy of less than $\sim 1\%$ in the $FWHM$. For the
two filters we derived a resolution limit of about $0\farcs 6$ at 95\% confidence level.

As measurements of point-like sources we have chosen for $LW3$ and $LW10$ filters observations of HI78527 and of HIC80331, respectively.
As in the case of SN~1987A the observation provides a sampling of the source with an angular resolution of $1\arcsec$ with a pixel size of $3\arcsec$ (Tab. \ref{starmeas}).
The data were calibrated in an almost identical way as the observation of SN~1987A
with the $LW2$ filter. The only difference was that
the deglitching for the calibration star data could be done fully automatically using standard procedures due
to the benign radiation environment encountered for these observations.
From the data we first derived four images with a sampling of $1\arcsec$ and calculated the final one as a weighted
average.

\begin{table}
  \caption{\label{starmeas}
    Observation log of the calibration stars}
  \begin{tabular}{l||c|c}
    Filter & $LW10$ & $LW3$ \\
    \hline
    target & HIC78527 & HIC80331 \\
    TDT    &  189007 &  119036  \\
    observer & CAM CAL & CAM CAL\\
    observation date  & 15.3.1996 & 24.5.1996\\
    raster & $6\times 6$ & $6\times 6$ \\
    pixel size & $3\arcsec$ & $3\arcsec$ \\
    step size & $2\arcsec$ & $2\arcsec$ \\
    read out interval & 0.28 s & 0.28 s\\
    time between pointings & 60 s & 60 s\\
  \end{tabular}
\end{table}

\subsubsection{Extension measurements for the calibration stars}

\label{test1}
The probability distributions in source size of the calibration stars were
derived from background-subtracted images using
theoretical $psf$s for a
spectral index of $k=4$ appropriate for the Rayleigh-Jeans part of the spectrum.
For the calculation we used a subimage of $11\times 11$ pixels centred on the source.

\begin{figure*}
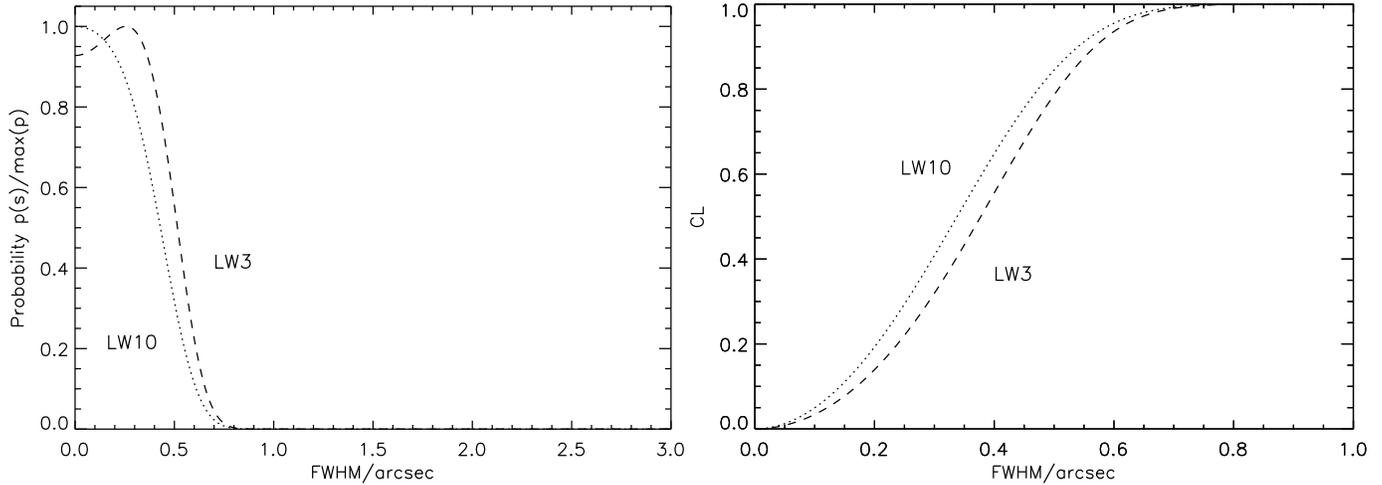

  \resizebox{0.5\hsize}{!}{\includegraphics{ms1827f11.epsi}}
  \resizebox{0.5\hsize}{!}{\includegraphics{ms1827f12.epsi}}
  \caption{\label{result_star}
    Result of the size measurement for the point-like sources HIC80331 ($LW3$-filter) and HIC78527 ($LW10$-filter).
    The derived probability distributions of the size $s$ (see Eq.~\ref{source_s})
    of the theoretical image $B$ (Eq.~\ref{imagefkt}) as a function
    of the $FWHM$ of the Gaussian source is shown in the left panel. The right panel
    gives the corresponding confidence levels for upper limits on source size derived using Eq.~\ref{conflevel}.}
\end{figure*}

The result
is shown in Fig. \ref{result_star}. 
As for the supernova data we have taken the uncertainties of the pixel values to be equal and
assigned a value such that the minimum $\chi^2$ corresponds to a confidence level of $90~\%$. 
The most probable sizes are found to be much smaller than the $FWHM$ of the $psf$.
For both filters we found that the source is smaller than $\sim 0\farcs 6$ with a confidence level of $95\%$. This is consistent with
the rough estimate of $0\farcs 6$ given in Sect.~\ref{accuracy}.

\subsubsection{Comparison of the size measurements of the MIR emission using observed and theoretical $psf$s}

\label{test2}
No point-like calibration source with the same spectrum of the source associated with SN~1987A was observed
with ISOCAM. However, the effect of interchanging the theoretical $psf$ with the observed $psf$
could still be evaluated. This was done by comparing calculations for the size probability distributions
at $12~\mu{\rm m}$, respectively
using as $psf$ the observed image of a star and a theoretical $psf$ for a spectral index $k=4$. 
To achieve
for the image of the calibration star the same sampling of $1/3\arcsec$ as for the theoretical $psf$
we used a cubic interpolation applied to the logarithm of the pixel values.
It can be seen from Fig.~\ref{sizeprobdistr} that a most probable size of $1\farcs 9$ is derived both for the measured
and the theoretical $psf$.
The larger size derived by using the measured image of a calibration star instead
of the theoretical $psf$ appropriate for the supernova is caused by 
the steeper spectrum.



\section{Position of the reference star}

\label{starposition}
To determine the position of the IR-emitting star we used archival observations with the WFPC2 of the HST,
made in V-band in February 1996
and three observations (two in R one in V-band) made in 1995 of the supernova with the SUSI-instrument 
(\emph{SUperb-Seeing Imager}) of the NTT (\emph{New Technology Telescope}). 
The V-band filter was also used to determine the correction 
of the distortion on the
WFPC2-fields (Holtzman et al., \cite{Holtzman95}).
The measurement with the NTT
was necessary due to the fact that the supernova and the IR-emitting star were on different CCDs
of the observation with the WFPC2, which can give an additional error in the astrometry. The astrometry
on one single CCD of the WFPC2 is accurate to $0\farcs 1$.
The only object, whose coordinates are determined with high accuracy is the SN~1987A itself.
To remove the offset in the astrometry 
of the WFPC2 data, 
we have taken the position 
$\mathrm{\alpha=5h35m27.968s,\,\delta=-69^{\circ}16'11\farcs 09}$ (J2000)
from Reynolds et al. (\cite{Reynolds95}),
which is accurate to $\sim 70\times 10^{-3}$ arcsec.
We removed the distortion on the WFPC2-field using the
polynomial correction given by Holtzman et al. (\cite{Holtzman95}).

To derive the astrometry of the observations made with the SUSI instrument, we projected
all possible star positions, seen on a single CCD of the WFPC2, with a tangential projection on the
SUSI field and compared these coordinates $(x_i,y_i)$ with the star positions $(\tilde x_i,\tilde y_i)$
on the SUSI field with a non linear $\chi^2$-fit:
\begin{equation}
  \chi^2=\sum_i\frac{\left((x_i-\tilde x_i)^2+(y_i-\tilde y_i)^2\right)^2}{\sigma^2}.
\end{equation}
The positions of the individual stars we derived fitting a two dimensional symmetrical Gauss function
to the single stars in the fields.

We found no indication of a distortion on the SUSI field.
The final derivations of the star positions in the observations with the SUSI instrument and the 
WFPC2 were accurate by $\sigma=0\farcs 025$ in the case
of the PC-field of the WFPC2 and by $\sigma=0\farcs 049$ in the case of the CCD with
the star (WF3). This is consistent with the accuracy given for the single CCDs of the WFPC2 after
correcting for distortion ($\sigma\approx 0\farcs 04$, Holtzman et al., \cite{Holtzman95}). 
The single CCDs indeed show a small offset of up to $\sim 0\farcs 13$. On the other hand we found
no measurable offset for the two CCDs with the supernova and the single IR-emitting star.
The absolute coordinates of the single IR-emitting star in the SUSI field are found to be 
\begin{equation}
  \alpha=5{\rm h}~35{\rm m}~18.42{\rm s},\quad\delta=-69^{\circ}~16'~30\farcs 65.
\end{equation}
The coordinates directly derived from the WFPC2 observation only
differed by $\sim 0\farcs 04$ in right ascension and by $\sim 0\farcs 05$ in declination.
The position of the star relative to the SN~1987A should be accurate to $\sim 0\farcs 05$. The uncertainty
of the absolute coordinates should be $\sim 0\farcs 09$.

\end{document}